\documentclass{aastex}
\usepackage{emulateapj5}

\newcommand{\etal}{et~al.\ }

\newcommand{\feka}{\hbox{Fe\,K$\alpha$}}

\newcommand{\kms}{\hbox{km~s$^{-1}$}}
\newcommand{\cmsq}{\hbox{cm$^{-2}$}}

\newcommand{\flux}{\hbox{erg~cm$^{-2}$~s$^{-1}$}}
\newcommand{\lumin}{\hbox{erg~s$^{-1}$}}

\newcommand{\simgt}{\lower 2pt \hbox{$\, \buildrel {\scriptstyle >}\over {\scriptstyle\sim}\,$}}
\newcommand{\simlt}{\lower 2pt \hbox{$\, \buildrel {\scriptstyle <}\over {\scriptstyle\sim}\,$}}
\newcommand{\msun}{\hbox{${M}_{\odot}$}}

\newcommand{\nh}{\hbox{${N}_{\rm H}$}}
\newcommand{\cross}{QSO~2237+0305}

\newcommand{\clover}{H~1413+117}
\newcommand{\cxo}{{\emph{Chandra X-ray Observatory}}}
\newcommand{\asca}{{\emph{ASCA}}}
\newcommand{\chandra}{{\emph{Chandra}}}

\newcommand{\rosat}{{\emph{ROSAT}}}

\newcommand{\hst}{{\emph{HST}}}

\slugcomment{Accepted by ApJ}
\shorttitle{\emph{CHANDRA} OBSERVATIONS of \cross}
\shortauthors{DAI ET AL.}

\tighten

\begin{document}

\def\sarc{$^{\prime\prime}\!\!.$}
\def\arcsec{$^{\prime\prime}$}

\title{\emph{CHANDRA} Observations of \cross}

\author{X. Dai,\altaffilmark{1} G. Chartas,\altaffilmark{1} E. Agol,\altaffilmark{2,3} M. W. Bautz,\altaffilmark{4} and G. P. Garmire\altaffilmark{1}}

\altaffiltext{1}{Astronomy and Astrophysics Department, Pennsylvania State University,
University Park, PA 16802, xdai@astro.psu.edu, chartas@astro.psu.edu,
garmire@astro.psu.edu}

\altaffiltext{2}{Theoretical Astrophysics, Caltech, Pasadena, CA 91125, agol@tapir.caltech.edu}

\altaffiltext{3}{{\it Chandra} Fellow}

\altaffiltext{4}{MIT Center for Space Research, 70 Vassar Street, Cambridge, MA, 02139, mwb@space.mit.edu}

\begin{abstract}
We present the observations of the gravitationally lensed 
system \cross\ (Einstein Cross) performed with the Advanced CCD Imaging 
Spectrometer (ACIS) onboard the \cxo\ on 2000 September 6, and on 2001 December 8
for 30.3~ks and 9.5~ks, respectively.
Imaging analysis resolves the four X-ray images of the Einstein Cross.
A possible fifth image is detected; however,
the poor signal-to-noise ratio of this image combined with 
contamination produced by a nearby brighter image
make this detection less certain. We investigate possible
origins of the additional image.
Fits to the combined spectrum of all images of the Einstein Cross
assuming a simple power law with Galactic and intervening absorption at the lensing 
galaxy yields 
a photon index of $1.90^{+0.05}_{-0.05}$ consistent with the range of 
$\Gamma$ measured for large samples of radio-quiet quasars.
For the first \chandra\ observation of the Einstein Cross this spectral
model yields a 0.4--8.0~keV X-ray 
flux of $4.6\times10^{-13} \flux$ and a 0.4--8.0~keV lensed luminosity
 of $1.0\times10^{46} \lumin$.  
The source exhibits variability both over long and short time scales.
The X-ray flux has dropped by 20\% between the two observations, and the 
Kolmogorov-Smirnov test showed that image A is variable at the 97\% confidence
level within the first observation.
Furthermore, a possible time-delay of $2.7^{+0.5}_{-0.9}$~hours between images A and B with image A leading is detected in the first \chandra\ observation.
The X-ray flux ratios of the images are consistent with the optical flux 
ratios which are affected by microlensing suggesting 
that the X-ray emission is also microlensed.
A comparison between our measured column densities and those
inferred from extinction measurements
suggests a higher dust-to-gas ratio in the
lensing galaxy than the average value of our Galaxy.
Finally, we report the detection at the 99.99\% confidence level  
of a broad emission feature near the redshifted energy of the 
Fe K$\alpha$ line in only the spectrum of image A. 
The rest frame energy, width, and equivalent width of this feature are  
$E_{line}$ = $5.7_{-0.3}^{+0.2}$~keV,
$\sigma_{line}$ = $0.87_{-0.15}^{+0.30}$~keV, 
and $EW = 1200^{+300}_{-200}$~eV, respectively.
\end{abstract}


\section{Introduction}
\cross\ was discovered by \citet{h85} as part of the Center for Astrophysics 
galaxy redshift survey. Four images have been resolved \citep{s88,y88} 
from this lens system. 
The quasar is at a redshift of $z_{s} = 1.695$ and the lensing galaxy 
is at a redshift of $z_{l} = 0.0395$. 
Microlensing induced by stars in the lens galaxy was proposed for this system 
shortly after its discovery \citep{k86, k88, s88, k89}, and was first confirmed by \citet{i89}.
Microlensing in \cross\ has been firmly established
with extensive monitoring of this system over several years \citep{r92,o96,w00a,w00b}.
The presence of both macrolensing and microlensing in the Einstein Cross 
and the proximity of the lensing galaxy, an order of magnitude closer than 
other lensing galaxies, make it a unique laboratory to explore the 
structure of the different emission regions in the source quasar, 
and the properties of the lensing galaxy as well.

During the past decades, major progress has been made in understanding 
the physical processes associated with Active Galactic Nuclei (AGN);
however, it is beyond the capabilities of current telescopes 
to resolve directly the central parts of AGNs. 
Observations of quasar microlensing events provide a means of exploring the structure of AGNs.
Multi-wavelength studies of \cross\ \citep{f96,m98,a00} 
demonstrate that the flux ratios of the images are different in different bands, 
in particular, between the radio, C~{\sc iii]}, mid-infrared, and optical.
The differences of flux ratios in different wavelength bands has been
interpreted as the result of microlensing.
Emission regions with a size significantly
less than the Einstein radius of the microlens in the source plane 
will be significantly magnified, whereas, emission regions with a size significantly
larger than the Einstein radius  will not be affected by microlensing.
Studies of \cross\ indicate that the radio and mid-infrared emission regions are
larger than the Einstein radius, whereas, the optical emission regions  
are smaller than the Einstein radius.
The C~{\sc iii]} emission regions also extend beyond the Einstein radius 
but are less extended than the radio and mid-infrared regions.
The analysis of light-curves of microlensing events can also constrain 
the source size of individual emission regions, 
especially in a high magnification event (HME) that may occur
during a caustic crossing.
\citet{y01} analyzed the light-curves of \cross\ and
estimated a source size of less than 2000 AU for the optical 
emission region in \cross. It is desirable to study the X-ray emission 
of a microlensed quasar since the X-rays are thought to originate 
from the inner most region of the accretion disc.
The X-ray light-curves of the images and the profile of the \feka\ line during a 
microlensing event, especially from a HME, could constrain the X-ray emission 
region and possibly yield information about the mass and 
spin of the central black hole \citep{y98,a99,p02a}.
Recently, a microlensing event in X-rays was observed by \citet{c02} 
in MG~J0414+0534, where an enhancement of the equivalent width of 
the \feka\ line was observed in only one of the images.

As mentioned earlier the proximity of the lensing galaxy 
of 2237+0305 has facilitated its detailed study in several wave bands. 
In particular, 
\citet{y88} detected a $g - i$ color difference between different 
image components which indicates differential extinction.
The extinction curve for the lensing galaxy was measured by \citet{n91}.
\citet{f92} measured the central velocity dispersion of
the barred spiral galaxy 2237+0305 to be $215\pm30~\kms$.
Modeling of the lens system has also provided estimates of the mass 
distributions of the lensing galaxy. In particular, \citet{s98} studied 
the contribution of the galaxy bar to the lens potential and 
found a bar-mass of about $7.5\times10^{8}\msun$.

\cross\ was first detected in X-rays with \rosat\ by \citet{w99}; 
however, due to the low spatial resolution of \rosat, 
the individual images were not resolved. To obtain
spatially resolved X-ray spectra from the individual images we
performed  \chandra\ observations of \cross.  
Here, we present the results of these observations.
We use $H_{0}$ = 65 km s$^{-1}$ Mpc$^{-1}$,
$\Omega_{m} = 0.3$, and $\Omega_{\Lambda}$ = 0.7, unless mentioned otherwise.

\section{Observations and Data Reduction}
\cross\ was observed with ACIS
\citep{g02} onboard the \cxo\ for $\sim30.3$~ks and $\sim9.5$~ks 
on 2000 September 6, and 2001 December 8, respectively.  
The data were taken continuously with no interruptions within each observation.
\cross\ was placed at the aim point of the ACIS-S array which 
is on the back-illuminated S3 chip. The data were reduced with the CIAO 2.2 software 
tools provided by the \emph{Chandra X-ray Center} (CXC).
We improved the image quality of the data by removing the pixel 
randomization applied to the event positions
in the CXC processing and by applying a subpixel resolution 
technique \citep{t01,m01}.
In the first and second observations of \cross\  we detected background flares
with durations of $\sim10$~ks and $\sim0.2$~ks, respectively.
We did not remove events collected during the background flares
in either of the two observations because even at the peak of the flares, the backgrounds only contribute by about 0.3\% and 1\% of the counts of \cross\ for the first and second observation, respectively. In the data analysis, only events with 
standard \asca\ grades of 0, 2, 3, 4, and 6 were used.

\section{Photometry and Astrometry}
\label{sec:astro}
Totals of 2632 and 607 source events were detected from 
circles centered on the centroids of the sources with radii of 3\arcsec\ and within the 0.2-10 keV energy band 
during the first and second observations of \cross, 
respectively. 
We applied a point spread function (PSF) fitting method to estimate 
the X-ray count rates of individual images due to the closeness of the 
individual images.
We modeled the \chandra\ images of A, B , C and D with PSFs generated by the simulation tool \verb+MARX+ \citep{w97}.
The X-ray event locations were binned with a bin-size of 0\sarc0246.
The simulated PSFs were fit to the \chandra\ data by minimizing the
Cash $C$ statistic formed between the observed and simulated images
of \cross. 
The relative positions of the images were fixed to the observed \hst\ values
obtained by the CfA-Arizona Space Telescope LEns Survey (CASTLES).
The CASTLES website is located at \verb+http://cfa-www.harvard.edu/glensdata/+.
The total count rate and count rates of individual images for 
each observation are listed in Table~\ref{tab:countrate}.
The total count rate of all images for the second observation 
has dropped by $27\pm4\%$ compared to the first observation.
The count rate ratios of images B and D with respect to image A 
are consistent within $1 \sigma$ errors between the two observations, 
and the ratio of C/A has decreased by $16\pm11\%$ 
in the second observation compared to the first observation.

\begin{figure}
\plotone{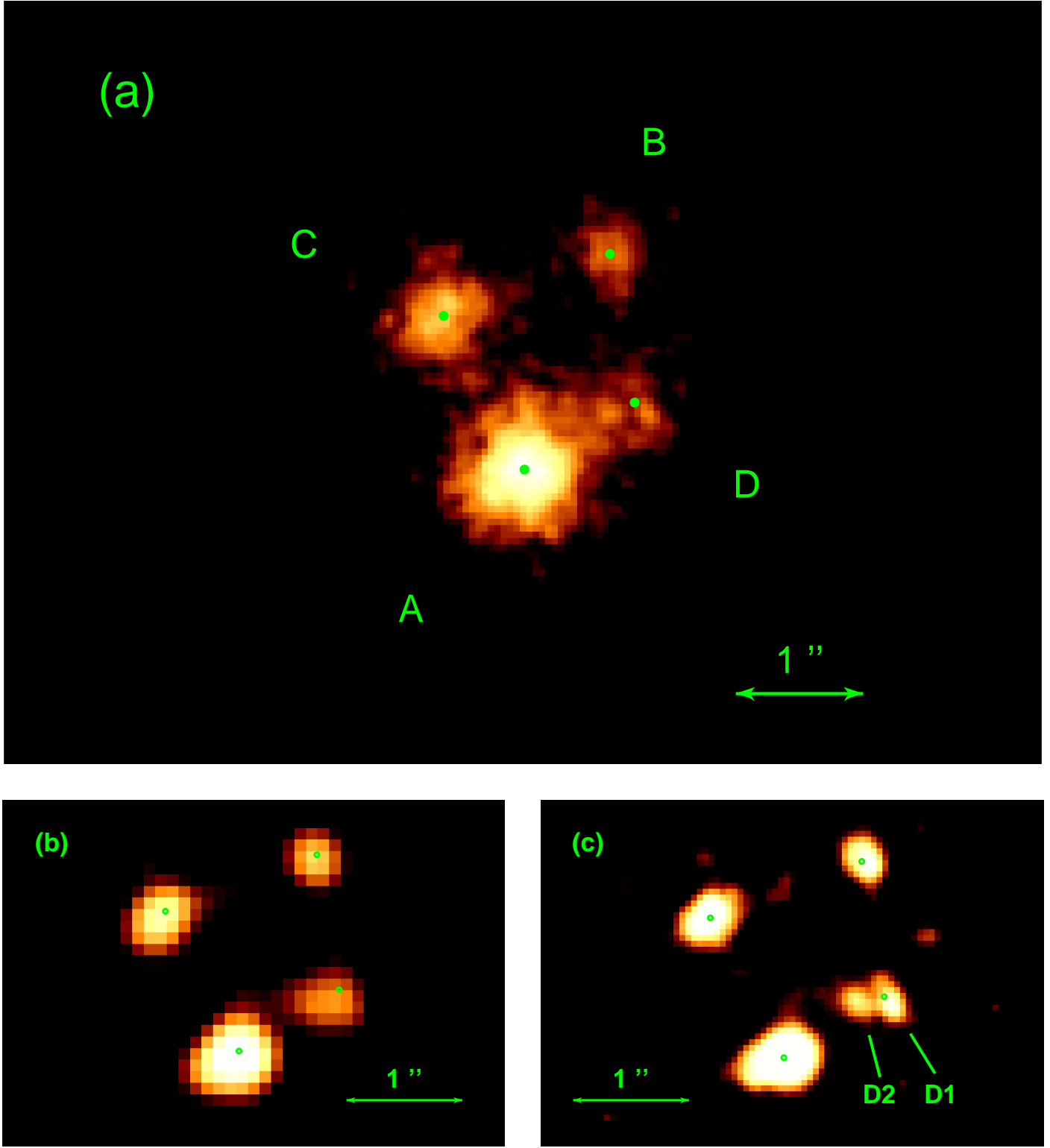}
\caption{(a) The raw image of the combined observations of \cross\, 
binned with a binsize of 0\sarc05 and smoothed with a Gaussian of 
0\sarc05 (top).  (b) The deconvolved image binned with a binsize of
0\sarc1 (bottom left).  (c) The deconvolved image binned with a binsize
of 0\sarc05 (bottom right).  
The green circles in each image are the corresponding \hst\ image 
positions provided by CASTLES (http://cfa-www.harvard.edu/glensdata/Individual/Q2237.html).  \label{fig:image}}
\end{figure}

The image of \cross\ obtained by combining the two 
\chandra\ observations is shown in Figure~\ref{fig:image}a.
The image is binned with a binsize of 0\sarc05 and smoothed with a Gaussian 
with $\sigma$ = 0\sarc05. Images B and C of \cross\
are clearly resolved in Figure~\ref{fig:image}a.
Images A and D are not well resolved, but since image A is the 
brightest image, it is less contaminated by image D.
Moreover, the raw binned image of D seems to have two concentrations 
of photons separated by about 0\sarc3.

The Lucy-Richardson deconvolution technique \citep{r72,l74} was applied 
to the combined image in order to resolve the four images.
We binned the X-ray events with binsizes of 0\sarc1 and 0\sarc05 for 
the deconvolution. The deconvolved images are 
shown in Figures~\ref{fig:image}b and \ref{fig:image}c.
In Figure~\ref{fig:image}b (image binsizes of 0\sarc1), a total of four images 
are resolved, which is consistent with previous observations in other wave bands, 
while in Figure~\ref{fig:image}c (image binsizes of 0\sarc05), five images in total are resolved.

The relative X-ray positions of the different images with respect 
to image A are obtained from the centroids of the deconvolved images.
X-ray image positions are listed in 
Table~\ref{tab:astrometry} together with the \hst\ positions 
obtained by CASTLES.
The relative positions of  images B and C with respect to A as measured 
with the deconvolution using a 0\sarc1 binsize are consistent 
with the \hst\ positions to 0\sarc03.
The separations between image D and the remaining images,
A, B, and C as measured with the deconvolution using a 0\sarc1 binsize
differ from those measured in the optical by 0\sarc09, 
0\sarc06, and 0\sarc07, respectively.
The expected observational uncertainties of the image
positions are $\sigma = \alpha/\sqrt{S/N}$, where $\alpha$ is the 
spatial resolution of ACIS and S/N is the signal-to-noise ratio of the image.
By accounting for the contamination of image D by image A , we estimate 
the uncertainty in the position of image D to be about 0\sarc1.
Thus, the discrepancy between the 
optical and X-ray positions of image D relative to image A 
 is about $1 \sigma$.

\section{Spectral Analysis}
Spectral analysis was carried out with the software tool \verb+XSPEC V11.2+ 
\citep{a96}. The total spectrum of all images of \cross\ was 
extracted from a circle 
of 3\arcsec\ radius centered on the centroid of the images. 
The spectra of images A, B, and C were extracted from circles of 1\arcsec\ 
radii centered on the images, and the spectrum of image D was extracted from a
circle of 0.6\arcsec\ radius centered between D1 and D2 in order to 
avoid the contamination from the brightest image A.
The background was extracted from an annulus centered on the centroid 
of the images with inner and outer radii of 5\arcsec\ and 30\arcsec, respectively.
All spectra were fitted in the 0.4--8 keV energy range assuming
Galactic absorption of $\nh = 5.5\times10^{20}\cmsq$ 
\citep{d90}. We have applied a correction to the ancillary response files 
to account for the use of relatively small extraction regions.
We determined the corrections to the ancillary response files by simulating 
the spectra of point sources at the locations of the images with and without 
apertures used in our analysis. For our simulations we used \verb+XSPEC+ to 
generate the source spectra 
and the raytrace tool \verb+MARX+ to  model the 
dependence of photon scattering with energy.
To account for the recently observed quantum efficiency decay of ACIS, 
possibly caused by molecular contamination of the ACIS filters, 
we have applied a time-dependent correction to the
ACIS quantum efficiency implemented in the \verb+XSPEC+ model \verb+ACISABS1.1+.
\verb+ACISABS+ is an \verb+XSPEC+ model contributed to the \chandra\
users software exchange web-site  
\verb+http://asc.harvard.edu/cgi-gen/cont-soft/sof+
\verb+t-list.cgi+. 
The ACIS quantum efficiency decay is insignificant for energies above 1~keV
and does not affect the main results of our analysis.

\subsection{Simple Absorbed Power-law Models}
\label{sec:pow}
\begin{figure}
\plotone{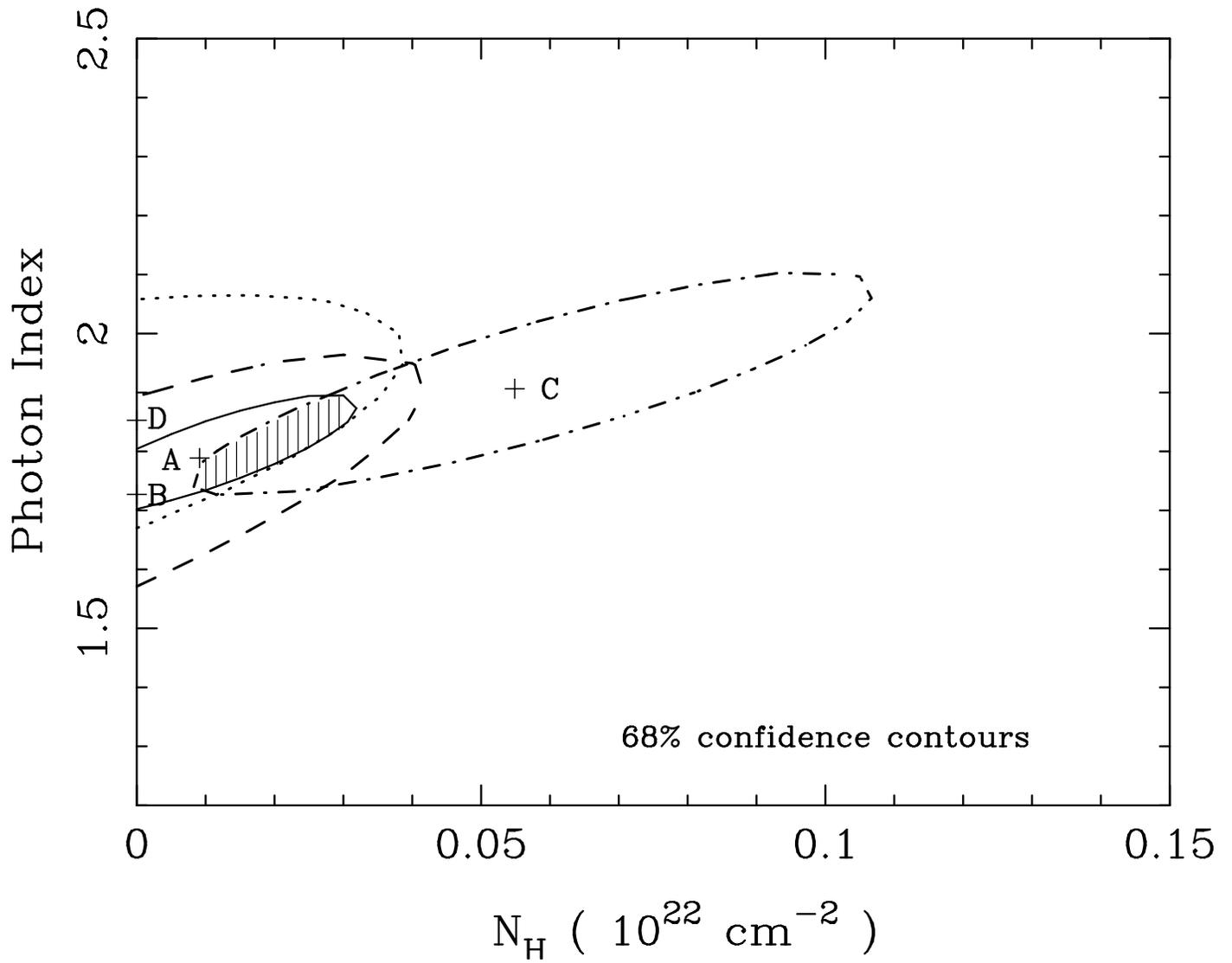}
\protect\caption{68\% confidence contours of photon index and neutral 
absorption at the lens ($z=0.0395$) for images A, B, C, and D of the 
combined spectra from both observations.  The shaded region represents
the overlapping region of the four confidence contours.
\label{fig:contour}
}
\end{figure}

We began by fitting the spectra of the individual images and the spectrum of all 
images  of \cross\  from each epoch with 
a simple power law modified by Galactic absorption and 
neutral absorption placed at the redshift of the lens ($z=0.0395$).  
To obtain tighter constraints on the model parameters we
fitted the spectra from both epochs simultaneously.  
In the case of the simultaneous fits, the model parameters, 
$\Gamma$ and \nh\ were kept the same between observations,
whereas, the normalization parameters were allowed to vary independently.
We also followed a different approach of fitting the combined spectra from both epochs.
In the case of the combined spectra we
weighted the ancillary response files from each observation.
The spectral fitting results are listed in Table~\ref{tab:spec}.
The simultaneous fits (fits 7-11 of Table 3) and the combined fits (fits 12-16 of Table 3) to
the spectra  of \cross\  yield consistent results.  We will use the results from the
combined spectra of both epochs in the remaining analysis  
since they provide slightly tighter constraints. 
The fit to the spectrum of all images for the combined observations 
yields a photon index of $\Gamma=1.90^{+0.05}_{-0.05}$ and a column 
density of $\nh=0.02^{+0.01}_{-0.01}\times10^{22}\cmsq$.
The column density is relatively low compared to that detected in 
other galaxies.
In Figure~\ref{fig:contour} we show 
the 68\% confidence contours of \nh\ versus photon indices
for all images. 
Absorption from gas in the lensing galaxy is marginally detected (at the 68\% confidence level)
towards image C.  For the lines of sight towards the remaining  images we can only place upper
 limits on the neutral hydrogen column densities from the lensing galaxy.
In  Table~\ref{tab:spec} we also list the 0.4--8~keV X-ray fluxes for fits 
1---6. 
The fluxes of individual images were normalized based on the PSF fitting 
results in section 3.
The total 0.4--8~keV fluxes for the first and second observations 
of \cross\ are $4.6^{+0.4}_{-0.2}\times10^{-13}\flux$, and $3.7^{+0.6}_{-0.5}\times10^{-13}\flux$, 
and 0.4--8.0~keV lensed luminosities are $1.0\times10^{46}\lumin$, and $8.3\times10^{45}\lumin$, respectively.
These luminosities have to be corrected by the lensing magnification
in order to obtain the true luminosity of the quasar.
The range of macro magnification was estimated from a few up to many hundred \citep{w94}, and a recent model from \citet{s98} estimated the magnification to be $16^{+5}_{-4}$.

\subsection{Features in the Spectrum of Image A}
\begin{figure}
\epsscale{0.9}
\plotone{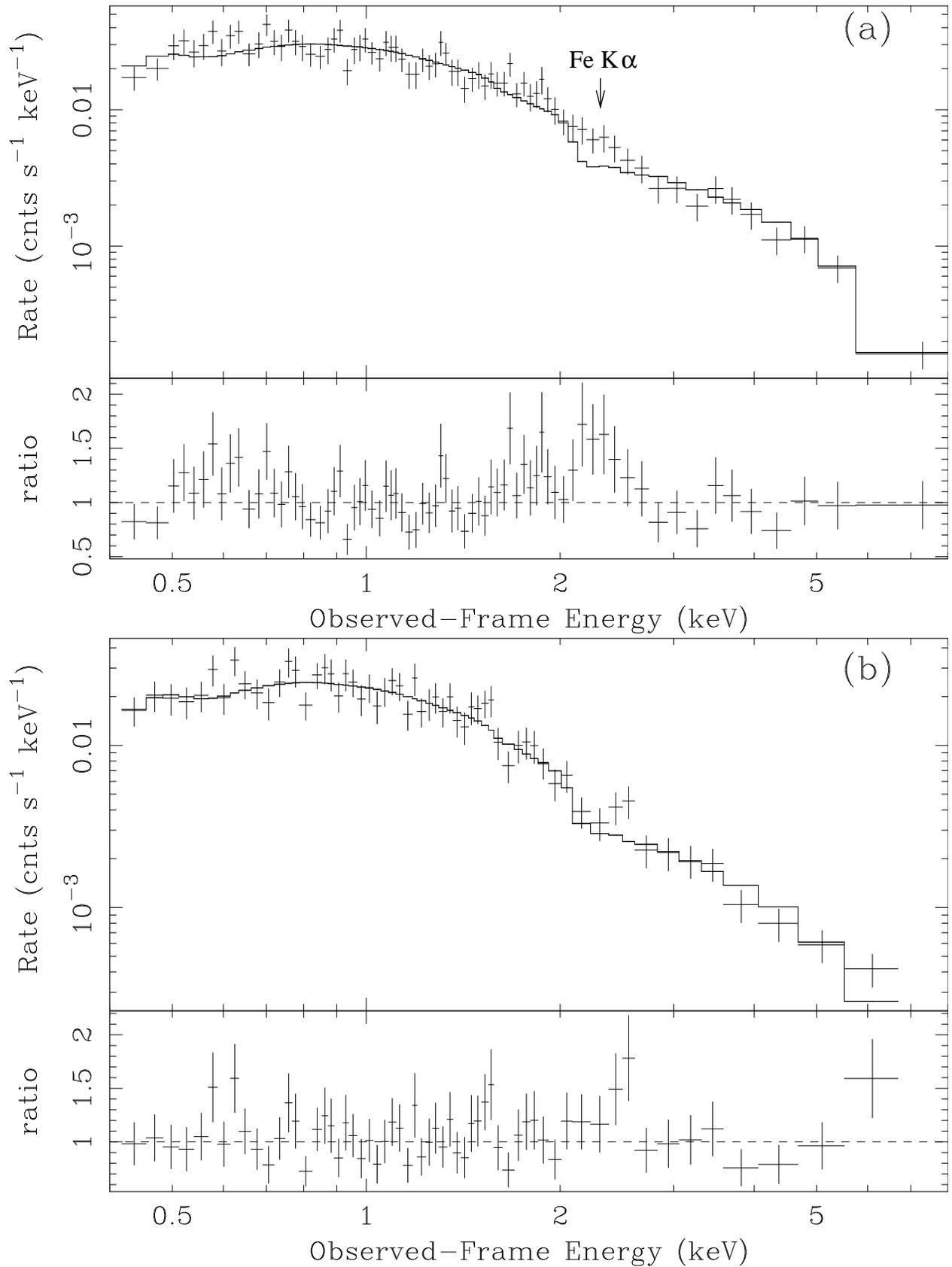}
\protect\caption{(a) Spectrum of image A combined from both observations (top).  
(b) Combined spectrum of images B, C, and D from both observations (bottom)\label{fig:spectra2}.
}
\end{figure}

Figure~\ref{fig:spectra2}a shows the spectrum of image A combined
from both observations of \cross\ overplotted with a
simple absorbed power law model described in $\S$\ref{sec:pow}. 
The spectrum shows residuals in the 2--3 keV band, and the residuals
are near the red-shifted energy of the \feka\ line.
For a comparison in Figure~\ref{fig:spectra2}b we show the combined
spectrum of images B, C, and D for both observations overplotted
with the best fit spectral model. 
In the combined spectrum of B, C, and D we detect
residuals in the 2--3 keV band.
However, these residuals are not 
as significant as those in image A and the peaks of the residuals 
differ between the two spectra.
To model the residuals in image A 
we added a redshifted Gaussian line component 
to the absorbed power-law model and
present the best fit parameters in Table~\ref{tab:iron}.
The line energy and width are kept free parameters during the fitting.
Including a Gaussian emission line in our model for the spectrum
of image A leads to a significant
improvement in fit quality at the 99.99\% confidence level (according to the $F$-test).
The best-fit rest-frame energy, width, and equivalent width of the modeled emission line in the spectrum of
image A are $E_{line}$ = $5.7_{-0.3}^{+0.2}$~keV,
$\sigma_{line}$ = $0.87_{-0.15}^{+0.30}$~keV,
and $EW = 1200^{+300}_{-200}$~eV, respectively.
We also modeled the combined spectrum of images B, C, and D with a redshifted Gaussian component.
The improvement of the fit is significant at the 94\% level based on the $F$-test.
The rest-frame energy of the modeled Gaussian line of the 
combined spectrum of images B, C, and D is $E_{line}$ = $6.6_{-0.2}^{+0.1}$~keV
and its width is consistent with that of a narrow line.

\begin{figure}
\plotone{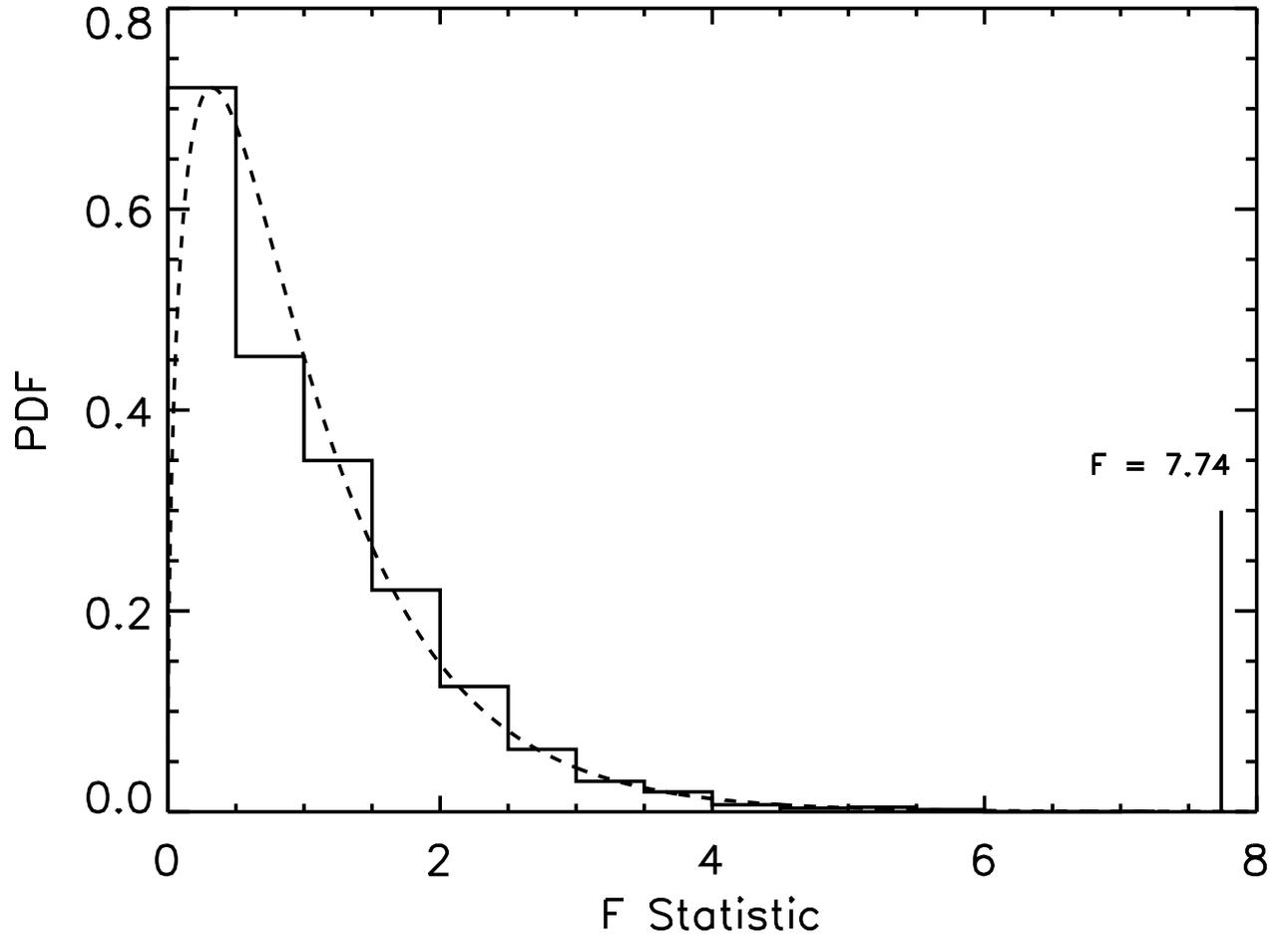}
\protect\caption{To evaluate the significance of the emission feature in image A, we determined the distribution of the F-statistic (shown as histogram) using a Monte-Carlo simulation.  We also overplotted for comparison the analytical curve for the F-distribution (shown as dashed curve).  The vertical line indicates the F value obtained from the real spectrum of image A.  The simulation shows that the detection of the broad emission line is 
significant at the $ > $ 99.99\% confidence level. \label{fig:monte}}
\end{figure}

\citet{p02} argued that the F-test cannot be applied to assess the significance of a line component in a spectral model because when the null values of the additional parameters (in our case these are the 
parameters of  the Gaussian line) fall on the boundary of the allowable parameter space, the 
reference distribution does not follow a F-distribution in general.
\citet{p02} proposed the method of posterior predictive p-value, a Monte-Carlo simulation approach, to calibrate the sample distribution of the F-statistic.
We followed this approach to calibrate the sample distribution of the F-statistic for the case of the spectrum of image A combined from both observations of \cross.
The parameters of the null model (powerlaw with Galactic absorption and absorption at the lens) are well constrained from the spectrum of $\sim1800$ photons and the simple approach described in section 5.2 of \citet{p02} is used.
We simulated 10,000 spectra with \verb+XSPEC+  from the null model with parameters fixed at their best fit value.   Each simulated spectrum, was binned as the real data and fitted with the null model and the alternative model 
(which included  an additional line component). The F-statistic between null and the alternative model 
was calculated for each simulation.
The results of the Monte-Carlo simulation are displayed in Figure~\ref{fig:monte}.
The maximum value of the F-statistics from the 10,000 simulated spectra is 7.28.
Therefore,  the F-statistic value of 7.74 obtained from the real spectrum of image A
indicates that the null model (which does not include an emission line) 
is rejected at the $  >  $ 99.99\% confidence level.
In addition, we also compared the Monte-Carlo simulated sample distribution with the analytical F-distribution in Figure~\ref{fig:monte}. The two distributions are  consistent in this particular case.

\section{Discussion}
Our spatial analysis of the \chandra\ observations of \cross\ 
has resolved the four lensed images A, B, C, and D, with positions in good agreement
with those obtained from \hst\ observations of \cross, however, both the raw and deconvolved
\chandra\ data of \cross\ hint to the presence of a fifth image near image D.
In section 5.1 we investigate plausible mechanisms that can explain the fifth image.
The analysis of the individual spectra of the images of \cross\ 
indicates the presence of absorption from the lensing galaxy.
In section 5.2 we compare the column densities obtained from our
X-ray analysis with the column densities inferred from the $g - i$
color changes of the images. This comparison is used to estimate the 
dust-to-gas ratio in the lensing galaxy.
In section 5.3 we compare the optical and X-ray image flux ratios.
An exciting finding of our spectral analysis was the identification
of a broad emission line near the energy of the redshifted Fe K$\alpha$
line in only image A.  In section 4.2 we performed Monte Carlo simulations
to show that the line in image A is significant at the 99.99\% level.
In section 5.4 we rule out possible instrumental effects that could mimic
such a line and discuss possible origins of the broad Fe K$\alpha$ line.
We discuss the variability of the source in section 5.5


\subsection{An Additional Image?}
\label{sec:adimg}
The \chandra\ observations of \cross\ indicate an additional 
image when the data is binned with a binsize of less than 0\sarc07 
both in the raw image and in the deconvolved image.
The total number of photons detected in both images D1 and D2 combining 
the two observations of \cross\ is $\sim340$, thus the apparent image splitting 
could be the result of poor photon statistics. 
The difference between the optical and
X-ray positions of image D is larger than that measured in the
other images. This large difference in image D may be the result of the 
significant contamination of this image by the brighter image A.
Considering the 0\sarc3 separation of D1 and D2, 
their X-ray fluxes, and the fact that the additional images only 
show up in the X-ray band, it would be extremely difficult 
to interpret this result, if the images D1 and D2 are real.
Here we briefly investigate possible origins (other than the mentioned statistical
interpretation) for the discrepancy between 
the optical and X-ray image configuration of \cross. 

(a) Spatially distinct X-ray flares in the source plane.\\
X-rays are generated in the inner most region of the accretion
disc. X-ray variability studies of quasars indicate
that the size of the X-ray continuum emission region is of
the order of $\sim$ 1 $\times$ 10$^{-4}$ pc (e.g., Chartas et~al.\ 2001). 
If we were to attribute the additional X-ray image to macrolensing or microlensing of 
two distinct X-ray flares the implied distance
between flares would be about 2 kpc in the source plane,
considerably larger than the expected size of the X-ray emitting region.
The flares would also have to vary over time-scales shorter
than the time-delay between the images since only one additional image is observed 
if we attribute it to macrolensing.

(b) An additional X-ray source in the lens plane.\\
The luminosity of the additional image would be $\sim10^{41}\lumin$
if the source producing the image were located in the 
lensing galaxy, and this is over two orders
of magnitude brighter than the Eddington limit for X-ray binaries.
This luminosity, however, lies at the upper end of the luminosity range ($10^{39-41}\lumin$, e.g., Roberts \etal 2002) of ultraluminous X-ray (ULX) sources. 

(c) An X-ray jet component extending from image A.\\
This is unlikely based on VLA observations that indicate \citep{f96} \cross\
being a radio quiet quasar with no jet.

(d) A source in our Galaxy.\\
The X-ray luminosity of the additional image would be $\sim10^{36}\lumin$
if it were located in the Galaxy, and any object with such a large X-ray luminosity
would have been detected by \hst.

(e) A fifth image produced by the lensing galaxy.\\
In principle, a gravitational macrolens could generate an odd
number of images if the lens potential is non-singular.
But the anticipated position of the central image should lie within the
central core of the lens potential and be greatly demagnified,
which is not consistent with the \chandra\ observations of \cross.

\begin{sloppypar}
(f) Microlensing by solar-mass stars in the central bulge of the lensing galaxy.\\
The image separation produced by a microlens of mass M is given by the expression
\hbox{$\alpha_{0}\approx3\times10^{-3}\sqrt{M/\msun}(D_{OL}/kpc)^{1/2}$~arcsec} \citep{s92},
where $D_{OL}$ is the angular diameter distance between observer and lens.
The interpretation of the 0\sarc3 image splitting of image D as 
a microlensing event would require a microlens mass of 
at least $\sim10^{8}\msun$.
The Einstein radius of the microlens at the source plane is proportional 
to the square-root of the mass of the microlens, specifically for this system, 
$R_{E} = 1.1\times10^{17}~M^{1/2}h^{-1/2}$~cm.
The optical and radio images should also show this splitting if such a large ``microlens'' 
were present, unless the optical and radio emission regions are far 
more extended than the Einstein radius.
This could possibly be true for the radio emission, but studies of 
microlensing of \cross\ showed that the optical emission region 
is within 1~pc of the accretion disc of the central engine 
\citep{y01}, which make the above interpretation highly improbable.
\end{sloppypar}

\subsection{Absorption at the Lensing Galaxy}
In the appendix of \citet{a00}, the extinctions for the 
images of \cross\ were estimated based on the $g-i$ color changes
between the images \citep{y88}, and the correlation between 
the $g-i$ color change and the lens galaxy surface 
brightness \citep{r91}.
The hydrogen column densities can be inferred from 
the extinction by assuming an extinction law of 
\hbox{$R = 3.1\pm0.9$} from the Milky Way and employing a 
dust-to-gas ratio of \nh= \hbox{$5.9\times10^{21}~E_{B-V}~mag^{-1}~cm^{-2}$} \citep{b78}.
A comparison between \nh\ obtained from our spectral 
fits to the \chandra\ data of \cross\ and the inferred values from the 
extinctions is presented in Table~\ref{tab:cden}.
The comparison shows that the column densities inferred from 
the extinction values are systematically larger than the 
column densities obtained from our X-ray analysis by about 
$2.1\sigma$, $2.5\sigma$, $2.6\sigma$, and $2.7\sigma$ 
for images C, B, A, and D, respectively.
The ratio of the column densities obtained from our X-ray analysis to those 
inferred  from the extinction values are about 20\%, 13\%, 6\%, and 5\% 
for image C, B, A, and D, respectively.
The estimated large value for \nh\ inferred from the extinction measurements
could arise from a problem with one of our two assumptions;
the extinction law and the gas-to-dust ratio taken to be that of the Milky Way.
The extinction law of the lensing galaxy 2237+0305 has been 
measured by \citet{n91} assuming \hbox{$A(K)/A(V) = 0.11$},
where A(K) and A(V) are the extinctions in the K band and V band, respectively.
They find that the resulting extinction law in \cross\ is in good agreement with 
that of the Milky Way. Thus, we are left with the possibility that the 
dust-to-gas ratio of the lensing galaxy 2237+0305 is significantly larger 
than that of the Milky Way.
The dust-to-gas ratios for several gravitational lenses are discussed by \citet{f99}.
They found that the dust-to-gas ratios are small for the systems B0218+357 and PKS~1830-211, which is opposite to what we observed in \cross.

\subsection{Image Flux Ratios}
It is well established from optical monitoring that \cross\ is being 
microlensed as mentioned in the introduction section. We compared the X-ray 
flux ratios of \cross\ with the optical V band flux ratios from the Optical 
Gravitational Lensing Experiment (OGLE)
monitoring data obtained only four days prior to our first observation.
An optical flux close to our second observation was not available.
The OGLE website is at \verb+http://bulge.princeton.edu/~ogle+.
The extinction for the optical data is corrected based on 
the method described in appendix of \citet{a00}.
To compute the X-ray flux ratios we used the 2--8~keV band to avoid 
the complication of absorption which may affect 
the soft energy band to a greater degree.
The results from this comparison are presented in Table~\ref{tab:fluxratio}.
The X-ray and optical V band flux ratios of images
B, C, and D relative to image A are consistent.
Since the optical fluxes are influenced by microlensing
the agreement between X-ray and optical V band flux ratios 
implies that the X-ray fluxes are also magnified
by microlensing.
We also listed the 2--8~keV flux ratios, with larger error bars though, 
for the second observation in Table~\ref{tab:fluxratio}.

\subsection{Broad Fe K$\alpha$ Line}
Our spectral analysis indicates the presence of significant residuals 
between energies of 2--3~keV in image A. 
This energy region is near the energy of the red-shifted 
Fe K$\alpha$ line. Unfortunately, these 
residuals fall near a sudden change of the HRMA/ACIS effective area
caused by the iridium M absorption edges of the \chandra\ mirrors. 
To ascertain any systematic calibration uncertainties near the mirror edge
we have fit the spectra of several test-sources\footnote{The test-sources observed with
ACIS S3 are the supernova remnant G21.5-0.9
observed on 2001 March 18, the millisecond pulsar J0437-4715
observed on 2000 May 29 and the radio-loud quasar Q0957+561 observed
on 2000 April 16.} with expected smooth power-law spectra
near the 2~keV iridium edge. Since $\chi^{2}$ residuals depend on 
the statistics, we filtered the test-source
data in time to produce spectra with a total
number of counts equal to that observed in the second
observation of \cross. Typical residuals for these
test-sources near the mirror edge
are less than $\sim$ 1$\sigma$ indicating that the observed
consecutive 2$\sigma$ residuals near 2~keV in the spectrum of image A 
are real and not
due to systematic errors in the calibration of the effective area of
the \chandra\ mirrors. 
We also note that the residuals are more significant in image A (at the 99.99\% confidence
level) than the combined spectrum of images B, C, and D (at the 94\% confidence level).
Furthermore, the energies and widths of the lines 
in image A and the remaining images differ significantly.
It is unlikely that the residuals are caused by pile-up.
We simulated, using the software tool \verb+LYNX+ (see appendix A of Chartas et~al.\ 2000),
a piled-up spectrum with input model 
parameters taken from fit 2 of Table 3 and using the observed count rate
in image A of 0.16 counts per frame.  
We find no evidence that the residuals are due to pile up.
To further test whether pile up is the cause of the observed
residuals, we excluded events from the core of image A.
After the removal of the core, the remaining spectrum of image A still
shows residuals between 2 and 3~keV.
We conclude that a broad Fe K$\alpha$ line is present in the 
spectrum of image A. 


The presence of a broad Fe K$\alpha$ line  in image A and not in the other 
images is indicative that microlensing may be enhancing the redshifted 
emission near the black hole. 
Observational and theoretical arguments rule out
interpretations other than microlensing.
First, strong and broad Fe K$\alpha$ lines detected 
in several low luminosity Seyfert galaxies are 
rarely observed in quasars.
This fact is further supported by the observed  anti-correlation
between the luminosities of AGN and the equivalent widths of  Fe K$\alpha$ lines detected in these AGN \citep{i93,n95} and by theoretical estimates of the properties of  Fe K$\alpha$ lines in quasars \citep{f00}. Recently, Fe K$\alpha$ lines with equivalent widths of $\sim$ 1~keV were
observed in the quasars  \clover\  \citep{o01,c03} and MG J0414+0534 \citep{c02}.  
However, these quasars are both 
gravitationally lensed and these relatively large equivalent widths 
were interpreted as the result of microlensing \citep{c02,p02a,c03}.
Second, an interpretation suggesting that the broad line in image A of \cross\ 
is being emitted by an X-ray flare which is only seen in image A is not supported by the \chandra\ observations.
In particular, the predicted time-delays between images A and B, A and C, and A and D are $\sim$ 2, 16, and 5~hours, respectively \citep{s98}.
If the flare duration is longer than a few days,
we would have detected a similar enhanced and broadened Fe K$\alpha$ line
in the spectrum of the combined images of B, C, and D, since the relative
time-delays are less than a day.
For a flare duration of less than $\sim~8.4$~hours
we would have detected significant variability of the strength and shape of 
the Fe K$\alpha$ line during the first observation.
We did not detect such variability during the first observation, and therefore,
rule out flares with durations of less than 8.4 hours as being responsible
for the broad Fe K$\alpha$ line detected in image A only.

In the following arguments we assume
that the difference between the spectrum of image A and the
combined spectrum of images B, C, and D is produced
by microlensing, which magnifies the inner part of the accretion disc.
The broadening and the observed redshift of the Fe K$\alpha$ line in the spectrum of image A 
imply that the emission originates near the center of the 
black hole where special and general relativistic effects are important.
The broad and redshifted profile of the  
Fe K$\alpha$ line and its large equivalent width also
indicate that the microlensing caustic should be located near the
broad Fe K$\alpha$ line emission region.
The broad Fe K$\alpha$ line 
emission region where special and general relativistic effects are important
is expected to be very small $\sim 10~r_{g}$ and the magnification 
of this relatively small emission region close to a fold caustic 
scales approximately as the inverse square root of its distance to the caustic.
This microlensing event is fundamentally different from those discussed in theoretical 
studies of optical and UV broad emission lines 
(e.g. Popovi$\acute{\rm{c}}$, Mediavilla, \& Mu$\tilde{\rm{n}}$oz 2001; Abajas \etal 2002)
in that the locations and sizes of the emission regions and the line 
broadening mechanisms are all different.
\citet{p02a} discussed the influence of microlensing on the shape of the 
Fe K$\alpha$ line in AGN. 
A qualitative comparison between the profile of the observed broad 
Fe K$\alpha$ line in image A and the theoretical predictions of \citet{p02a}
also indicates that the microlensing caustic should be located near
the broad Fe K$\alpha$ line emission region. 

The microlensing interpretation has to explain the following facts 
indicated by the comparison of the optical and X-ray observations of \cross.
First, the OGLE photometric data show no peak in the V band light-curve 
of image A during the September 2000 observation.  
The V band light-curve of image A shows a 
microlensing event that peaked about ten months earlier than the first \chandra\ observation.
Delays between the peaks of the X-ray and optical light-curves
have been simulated to occur in caustic crossings  
in the case where the X-ray and optical emission regions differ significantly
 \citep{mi01}.
As we mentioned earlier the estimated size of the X-ray broad Fe K$\alpha$ line region
where special and general relativistic effects would produce the
observed distortion of the Fe  K$\alpha$ line is $\sim$ 10~$r_{g}$,
whereas, the size of the optical and X-ray continuum emission regions 
are 
$\simlt 6 \times 10^{15}$~cm and $\sim 1 \times 10^{14}$~cm, respectively.
The size of the X-ray continuum region is based on the
observed X-ray variability of the continuum 
in image A of about $\sim$~3000~sec (proper time)
which corresponds to a size of about $\sim 1 \times 10^{14}$~cm (see section 5.5
for more details).
The optical emission region has been constrained to be $\simlt$ 6 $\times$ 10$^{15}$~cm \citep{w00}. These sizes of emission regions are similar 
to those assumed in the simulations 
performed by \citet{mi01} that showed significant delays in the peaks of 
the magnification light-curves originating from different emission regions.
We note that the radiation mechanisms assumed for the emission regions
in the simulations of \citet{mi01} may differ from those in \cross, however, the main contribution to the simulated delays in the peaks arises from
the different sizes of the emission regions.
Second, the optical and X-ray continuum flux ratios are consistent within 
$1 \sigma$ at the time of the first \chandra\ observation.
This is possible since both the X-ray and optical continuum regions are much larger than 
the Fe K$\alpha$ line emission region, as we discussed above,
and they are both comparable in size with the Einstein radius 
of a typical 0.1 \msun\ microlens.
The simulations of \citet{mi01} also showed that the flux magnifications for the X-ray and optical extended emission regions are not significantly different during the microlensing event.
Detailed modeling of a caustic crossing that includes both the optical and X-ray constraints
of \cross\ are needed to accurately interpret the microlensing event in \cross.
However, such an analysis is beyond the scope of this paper.

\subsection{Variability}
We investigated the long term variability of \cross\ by comparing the fluxes 
of the two \chandra\ observations with the flux observed in the previous \rosat\ observation.
The 0.4--8~keV and 0.1--2.4~keV fluxes of \cross\ 
are $4.6\times10^{-13}\flux$ 
and $2.3\times10^{-13}\flux$, respectively, for the first \chandra\ observation.
The 0.1--2.4~keV flux detected with the first \chandra\ observation
is consistent with that previously detected with \rosat\ \citep{w99}.
A comparison between the two \chandra\ observations of \cross\
shows that the total 0.4--8~keV flux has decreased by about 20\%
in the second observation.
It is not clear from the present \chandra\ data if this decrease in
X-ray flux is caused by intrinsic variability of the quasar 
or by microlensing.

\begin{figure}
\epsscale{0.65}
\plotone{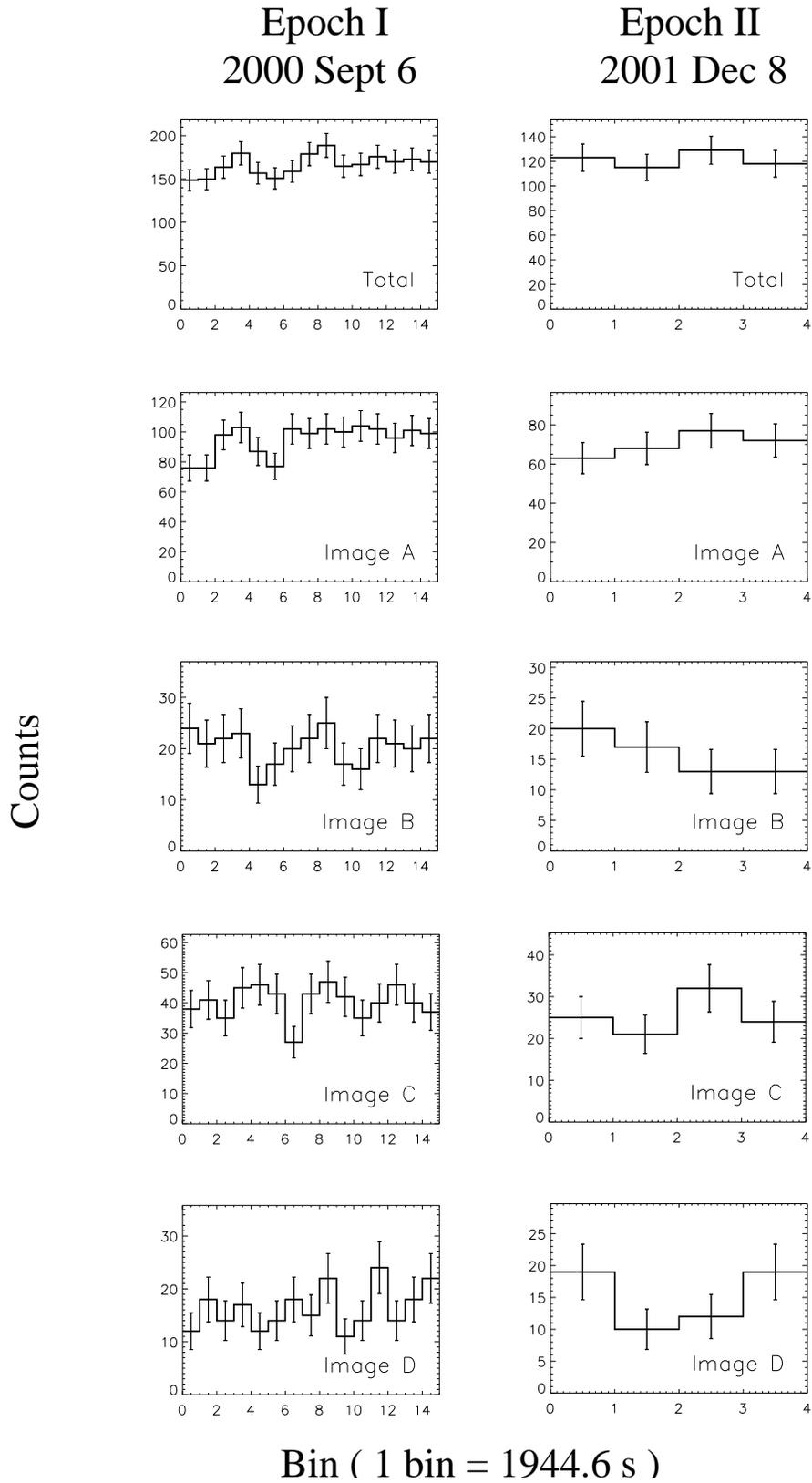}
\protect\caption{The full band (0.2--10 keV) light-curves of the sum of all images and images A, B, C, and D, separately, for the first and second \chandra\ observations of \cross.  The data were taken continuously within each observation.\label{fig:lca}}
\end{figure}

\begin{figure}
\plotone{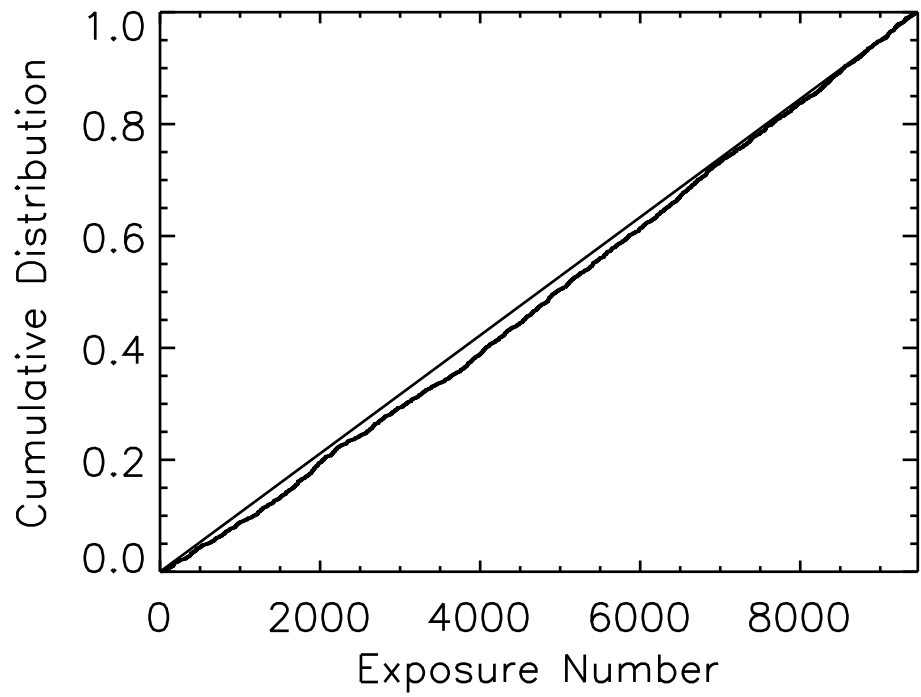}
\protect\caption{Cumulative probability distribution vs. exposure number for image A of the first observation of \cross\ compared to the cumulative probability distribution of a constant source. \label{fig:ksa}}
\end{figure}

We also explored the variability of \cross\ within each observation.
The light-curves of the combined images and the individual images A, B, C, and D for the first and second observation are displayed in Figure~\ref{fig:lca}.
We performed Kolmogorov-Smirnov (K-S) tests to the unbinned light-curves of each image.
The K-S test results are listed in Table~\ref{tab:ks}.
The K-S  results indicate that the light-curve of image A for the first observation is variable at the 97\% confidence level. 
The K-S plot of the cumulative probability distribution versus the exposure number for image A of the first observation is illustrated in Figure~\ref{fig:ksa}.
The light-curve of the combined images for the first observation also shows some variability according to the K-S test, however,  at a lower $\sim$ 90\% confidence level.
In this light-curve we identified two possible flux enhancements. 
A comparison between the light-curve of all images and that of image A indicated that the first enhancement in the total light-curve corresponds to the bump detected at a similar  time 
in the light-curve of image A.
The correspondence of the second flux enhancement of the total light-curve with bumps in individual light-curves is less clear than the first bump.  It is possible that the second bump 
arises in image B.
We performed an auto-correlation of the total light-curve for the first observation 
and obtained a lag of 5 bins (1 bin = 1944.6 seconds) which yields a maximum
auto-correlation coefficient of 0.40. The probability of obtaining this auto-correlation coefficient 
by chance is 0.06.
We tested the sensitivity of the computed lag-time to the selected bin size of the total
light-curve by performing the auto-correlation over a range of bin sizes. In all cases we recovered lag-times similar to the one obtained above. Our simple auto-correlation analysis indicated a possible time-delay of $2.7^{+0.5}_{-0.9}~hours$ in the total light-curve of \cross\ for the first \chandra\ observation.  The errors of this time-delay are dominated by the different choices of the bin size.  
A comparison between the total light-curve and those of the individual images indicates
that the measured time-delay most likely corresponds to the time-delay between images
A and B with image A leading. This is consistent with recent modeling of \citep{s98} that predict a
time-delay between images A and B of $2.0^{+0.4}_{-0.3}~h_{75}^{-1}$~hours with image A leading. 



\section{Conclusions}

\begin{enumerate}

\item {\chandra\ observations of \cross\ have resolved 
the system into at least four X-ray images.  
A possible fifth image could be the result of poor photon 
statistics and the contamination from the brightest image A.
A longer \chandra\ observation of \cross\ is needed to resolve this issue.}

\item {The X-ray flux ratios of images B, C, and D with respect to image 
A are consistent with the V band flux ratios observed only four days prior to our first \chandra\ observation.  
This indicates that the X-ray fluxes are also 
magnified by microlensing as expected since the 
X-rays are thought to originate from the inner most regions of the accretion disc.}

\item {The hydrogen column densities of images A, B, C, and D
measured from the \chandra\ observations of \cross\
are significantly lower than the column densities inferred from extinction measurements 
of these images in the optical and infrared bands.
This difference is suggestive of a higher value of
the dust-to-gas ratio in the lensing galaxy 
compared to the Galactic value.}

\item {Our spectral analysis indicates the presence of a broad Fe K$\alpha$
line in image A with a rest-frame energy, width, and equivalent width of 
$E_{line}$ = $5.7_{-0.3}^{+0.2}$~keV, 
$\sigma_{line}$ = $0.87_{-0.15}^{+0.30}$~keV
and $EW = 1200^{+300}_{-200}$~eV, respectively.
The enhancement of the emission line in image A 
is possibly caused by microlensing since the combined spectrum of the 
other three images does not show such a significant feature.
The redshift and broadening of the line may be the result of the
Doppler effect and special and general relativistic effects. }

\item {\cross\ exhibits variability both over long and short time scales from
the \chandra\ observations.
The X-ray flux has dropped by 20\% between the two observations, and the 
Kolmogorov-Smirnov test showed that image A is variable at the 97\% confidence
level within the first observation.  A possible time-delay of $2.7^{+0.5}_{-0.9}~hours$ between images A and B with image A leading is detected
in the first \chandra\ observation. 
}
\end{enumerate}

\acknowledgments
We would like to thank Pat Broos for providing the TARA software package, 
Koji Mori for providing subpixel correction software, 
and the anonymous referee for providing many useful comments and suggestions.
We acknowledge the financial support by NASA grant NAS 8-01128.

\clearpage

\begin{deluxetable}{ccccccccc}
\tabletypesize{\scriptsize}
\tablecolumns{9}
\tablewidth{0pt}
\tablecaption{\chandra\ Observations of \cross\tablenotemark{a}\label{tab:countrate}}
\tablehead{
\colhead{Observation} & 
\colhead{Exposure} & 
\colhead{Total\tablenotemark{b}} &
\colhead{$\rm{R_{Total}}$\tablenotemark{c}} & 
\colhead{$\rm{R_{A}}$\tablenotemark{d}} & 
\colhead{$\rm{R_{B}}$} & 
\colhead{$\rm{R_{C}}$} & 
\colhead{$\rm{R_{D}}$} & 
\colhead{$\rm{R_{Bkg}}$\tablenotemark{e}} \\
\cline{4-8}
\colhead{Date} & 
\colhead{$\rm{s}$} & 
\colhead{Counts} &
\multicolumn{5}{c}{$\rm{10^{-3}~cnts~s^{-1}}$} &
\colhead{$\rm{10^{-6}~cnts~('')^{-2}~s^{-1}}$} 
}

\startdata
2000-09-06 & 30287 & 2635 & $86.9\pm1.7$ & $47.3\pm1.3$ & $10.0\pm0.6$ & $19.0\pm0.8$ & $10.6\pm0.7$ & $4.5\pm0.2$ \\
2001-12-08 & 9538  & 608 & $63.6\pm2.6$ & $36.1\pm2.0$ & $7.3\pm0.9$  & $12.1\pm1.2$ & $8.1\pm1.0$ & $4.0\pm0.4$ \\
\enddata

\tablenotetext{a} {All events are of standard ASCA grade of 0,2,3,4,6 and have energies between 0.2--10~keV.}
\tablenotetext{b} {The total counts is the number of events including the background within a circle centered on the centroid of the source with a radius of 3''.}
\tablenotetext{c} {$\rm{R_{Total}}$ is the net count rate (background subtracted) of all images extracted from a circle centered on the centroid of the source with a radius of 3''.}
\tablenotetext{d}{$\rm{R_{A}}$,$\rm{R_{B}}$, $\rm{R_{C}}$, and $\rm{R_{D}}$ are the net count rates of images A, B, C, and D obtained after the PSF fitting, respectively.}
\tablenotetext{e}{$\rm{R_{Bkg}}$ is the count rate per $\rm{arcsec}^{2}$ of the background extracted from an annulus with inner and outer radii
of 5'' and 30'', respectively.}
\end{deluxetable}

\clearpage

\begin{deluxetable}{cccccccc}
\tabletypesize{\scriptsize}
\tablecolumns{8}
\tablewidth{0pt}
\tablecaption{Relative Optical and X-ray Positions of 
\cross\ Images\tablenotemark{a}\label{tab:astrometry}}
\tablehead{
\colhead{Telescope} & 
\colhead{Offset} &
\colhead{A} & 
\colhead{B} & 
\colhead{C} & 
\colhead{D} &
\colhead{D1} &
\colhead{D2} \\
\colhead{}  &
\colhead{}  &
\colhead{arcsec} &
\colhead{arcsec} & 
\colhead{arcsec} &
\colhead{arcsec} & 
\colhead{arcsec} &
\colhead{arcsec} 
}
\startdata
\hst\tablenotemark{b} & RA & 0 & $-0.673\pm0.003$ & $0.635\pm0.003$ & $-0.866\pm0.003$ & ... & ... \\
    & DEC & 0 & $1.697\pm0.003$  & $1.210\pm0.003$ & $0.528\pm0.003$  & ... & ... \\
\tableline
\chandra\tablenotemark{c} & RA  & 0 & $-0.67\pm0.02$ & $0.63\pm0.02$ & $-0.78\pm0.03$ & ... & ... \\
            & DEC & 0 & $1.73\pm0.02$  & $1.21\pm0.02$ & $0.49\pm0.03$  & ... & ... \\
\tableline
\chandra\tablenotemark{d} & RA  & 0 & $-0.68\pm0.02$ & $0.62\pm0.02$ & ... & $-0.93\pm0.04$ & $-0.67\pm0.04$ \\
            & DEC & 0 & $1.72\pm0.02$  & $1.21\pm0.02$ & ... & $0.46\pm0.04$  & $0.46\pm0.04$  \\
\enddata

\tablenotetext{a}{The image positions are relative to image A}
\tablenotetext{b}{The \hst\ positions are obtained from the CASTLES website (http://cfa-www.harvard.edu/glensdata/Individual/Q2237.html).}
\tablenotetext{c}{The \chandra\ positions are obtained from the 
centroids of the deconvolved image of \cross\ binned with a binsize of 0.1''.}
\tablenotetext{d}{The \chandra\ positions are obtained from 
the centroids of the deconvolved image of \cross\ binned with a binsize of 0.05''.}
\end{deluxetable}

\clearpage

\begin{deluxetable}{ccccccll}
\tabletypesize{\scriptsize}
\tablecolumns{8}
\tablewidth{0pt}
\tablecaption{Results of Fits to the \chandra\ Spectra
of \cross\tablenotemark{a}\label{tab:spec}}
\tablehead{
\colhead{Fit\tablenotemark{b}} & 
\colhead{Epoch} & 
\colhead{Image} & 
\colhead{$\Gamma$} & 
\colhead{$\rm{N_{H}}(z=0.0395)$} & 
\colhead{Flux\tablenotemark{c}} & 
\colhead{$\chi^{2}_{\nu}(\nu)$} & 
\colhead{$P(\chi^{2}/{\nu})$\tablenotemark{d}}
\\
\colhead{} &
\colhead{} &
\colhead{} &
\colhead{} &
\colhead{$10^{22}~\rm{cm^{-2}}$} &
\colhead{$10^{-13}~\rm{erg~s^{-1}~cm^{-2}}$} &
\colhead{} &
\colhead{}
}

\startdata
1 & I  & Total & $1.90^{+0.06}_{-0.05}$ & $0.02^{+0.01}_{-0.02}$ & $4.6^{+0.4}_{-0.2}$ & 1.22(124)  & 0.05 \\
2 & I  & A     & $1.81^{+0.07}_{-0.07}$ & $0.01^{+0.02}_{-0.01}$ & $2.5^{+0.2}_{-0.2}$ & 1.25(76)  & 0.07 \\
3 & I  & B     & $1.70^{+0.15}_{-0.12}$ & $0.00^{+0.03}_{-0.00}$ & $0.5^{+0.1}_{-0.1}$ & 1.11(15)   & 0.34 \\
4 & I  & C     & $1.93^{+0.07}_{-0.08}$ & $0.05^{+0.03}_{-0.04}$ & $1.0^{+0.1}_{-0.1}$ & 0.87(35)  & 0.68 \\
5 & I  & D     & $1.78^{+0.18}_{-0.12}$ & $0.00^{+0.03}_{-0.00}$ & $0.7^{+0.1}_{-0.1}$ & 0.75(12)   & 0.70 \\
\tableline 
6 & II & Total & $1.88^{+0.12}_{-0.12}$ & $0.03^{+0.03}_{-0.03}$ & $3.7^{+0.6}_{-0.5}$ & 1.26(33)  & 0.14 \\
\tableline
7 & I+II simultaneous & Total & $1.90^{+0.06}_{-0.06}$ & $0.02^{+0.01}_{-0.02}$ & ... & 1.22(159) & 0.03 \\
8 & I+II simultaneous & A     & $1.81^{+0.07}_{-0.07}$ & $0.01^{+0.02}_{-0.01}$ & ... & 1.31(96)  & 0.02 \\
9 & I+II simultaneous & B     & $1.69^{+0.14}_{-0.11}$ & $0.00^{+0.02}_{-0.00}$ & ... & 1.10(18)  & 0.34 \\
10 & I+II simultaneous & C    & $1.90^{+0.14}_{-0.14}$ & $0.04^{+0.04}_{-0.03}$ & ... & 0.83(41)  & 0.78 \\
11 & I+II simultaneous & D    & $1.83^{+0.16}_{-0.13}$ & $0.00^{+0.03}_{-0.00}$ & ... & 0.70(15)  & 0.78 \\
12 & I+II combined & Total & $1.90^{+0.05}_{-0.05}$ & $0.02^{+0.01}_{-0.01}$ & ... & 1.02(149) & 0.40 \\
13 & I+II combined & A     & $1.79^{+0.06}_{-0.07}$ & $0.01^{+0.01}_{-0.01}$ & ... & 1.06(94)  & 0.32 \\
14 & I+II combined & B     & $1.73^{+0.12}_{-0.10}$ & $0.00^{+0.02}_{-0.00}$ & ... & 1.26(20)  & 0.20 \\
15 & I+II combined & C    & $1.91^{+0.08}_{-0.07}$ & $0.05^{+0.03}_{-0.03}$ & ... & 0.89(41)  & 0.68 \\
16 & I+II combined & D    & $1.86^{+0.14}_{-0.13}$ & $0.00^{+0.02}_{-0.00}$ & ... & 1.13(16)  & 0.32 \\

\enddata

\tablenotetext{a}{All models include Galactic absorption with a column density of $\rm{N_{H}=0.055\times10^{22}~cm^{-2}}$.  All derived errors are at the 68\% confidence level.}
\tablenotetext{b}{The spectral fits were performed within the energy range 0.4--8 keV}
\tablenotetext{c}{Flux is estimated in the 0.4--8 keV band}
\tablenotetext{d} {$P(\chi^{2}/{\nu})$ is the probability of exceeding $\chi^{2}$ for ${\nu}$ degrees of freedom.}

\end{deluxetable}

\clearpage

\begin{deluxetable}{cccccccccll}
\tabletypesize{\scriptsize}
\tablecolumns{11}
\tablewidth{0pt}
\tablecaption{Fits to the Spectrum of Image A and the Combined Spectrum of 
Images B, C, and D of \cross\ with and without an Fe line\label{tab:iron} Model}
\tablehead{
\colhead{Fit} & 
\colhead{Epoch} & 
\colhead{Image} &
\colhead{Model\tablenotemark{a}} &
\colhead{$\Gamma$} & 
\colhead{$\rm{N_{H}}(z=0.0395)$} & 
\colhead{$\rm{E_{line}}$} & 
\colhead{$\sigma$} &
\colhead{EW} &
\colhead{$\chi^{2}_{\nu}(\nu)$} & 
\colhead{$P(\chi^{2}/{\nu})$\tablenotemark{b}} 
\\
\colhead{} &
\colhead{} &
\colhead{} &
\colhead{} &
\colhead{} &
\colhead{$10^{22}~\rm{cm^{-2}}$} &
\colhead{$\rm{keV}$} &
\colhead{$\rm{keV}$} &
\colhead{$\rm{eV}$} &
\colhead{} &
\colhead{} 
}

\startdata
1 & I+II & A   & pow          & $1.80^{+0.06}_{-0.07}$ & $0.01^{+0.01}_{-0.01}$ & ...                 & ...                    & ...  & 1.04(73) & 0.38        \\
2 & I+II & A   & pow+Gaussian & $1.83^{+0.07}_{-0.06}$ & $0.00^{+0.02}_{-0.00}$ & $5.7^{+0.2}_{-0.3}$           & $0.87^{+0.30}_{-0.15}$ & $1200^{+300}_{-200}$ & 0.82(70) & 0.87      \\
3 & I+II & B+C+D & pow          & $1.98^{+0.07}_{-0.13}$ & $0.03^{+0.02}_{-0.02}$ & ...                 & ...                    & ...  & 0.87(58) & 0.75     \\
4 & I+II & B+C+D & pow+Gaussian & $2.00^{+0.09}_{-0.08}$ & $0.03^{+0.02}_{-0.02}$ & $6.6^{+0.1}_{-0.2}$ & $< 0.30$      & $400^{+300}_{-200}$  & 0.81(55) & 0.85  \\  
\enddata

\tablenotetext{a} {All model fits include fixed, Galactic absorption of
$N_{\rm H}=5.5\times10^{20}$~cm$^{-2}$ (Dickey \& Lockman 1990).  The Fe line energy, width, and equivalent width are rest frame values ($z = 1.695$).}
\tablenotetext{b} {$P(\chi^{2}/{\nu})$ is the probability of exceeding $\chi^{2}$
for ${\nu}$ degrees of freedom.}
\end{deluxetable}

\clearpage

\begin{deluxetable}{ccccc}
\tabletypesize{\scriptsize}
\tablecolumns{5}
\tablewidth{0pt}
\tablecaption{Hydrogen Column Densities in \cross\ Images\label{tab:cden}}
\tablehead{
\colhead{} & 
\colhead{A} & 
\colhead{B} & 
\colhead{C} & 
\colhead{D} \\
\colhead{}  &
\colhead{\hbox{$10^{21}~cm^{-2}$}} & 
\colhead{\hbox{$10^{21}~cm^{-2}$}} &
\colhead{\hbox{$10^{21}~cm^{-2}$}} &
\colhead{\hbox{$10^{21}~cm^{-2}$}} 
}

\startdata
\chandra\tablenotemark{a} & $0.1^{+0.1}_{-0.1}$ & $0.0^{+0.2}_{-0.0}$ &  $0.5^{+0.3}_{-0.3}$ &  $0.0^{+0.2}_{-0.0}$ \\
Converted Value\tablenotemark{b} & $1.7^{+0.6}_{-0.6}$ & $1.6^{+0.6}_{-0.6}$ & $2.5^{+0.9}_{-0.9}$ & $2.2^{+0.8}_{-0.8}$ \\
\enddata

\tablenotetext{a}{Column densities are obtained from the spectral modeling of the \chandra\ observations}
\tablenotetext{b}{Column densities are converted from the extinction through an extinction law and a dust-to-gas ratio}

\end{deluxetable}

\clearpage

\begin{deluxetable}{cccccc}
\tabletypesize{\scriptsize}
\tablecolumns{6}
\tablewidth{0pt}
\tablecaption{X-ray and Optical Flux Ratio in \cross\label{tab:fluxratio}}
\tablehead{
\colhead{Band} & 
\colhead{Date} &
\colhead{A} & 
\colhead{B} & 
\colhead{C} & 
\colhead{D} 
}
\startdata
V\tablenotemark{a}     & 2000 Sep 2 & $ 1 $ & $0.24\pm0.07$ & $0.54\pm0.19$ & $0.29\pm0.09$ \\
X-Ray\tablenotemark{b} & 2000 Sep 6 & $ 1 $ & $0.23\pm0.05$ & $0.38\pm0.06$ & $0.27\pm0.06$ \\
X-Ray & 2001 Dec 8 & $ 1 $ & $0.27\pm0.19$ & $0.42\pm0.22$ & $0.19\pm0.10$ \\
\enddata

\tablenotetext{a}{The V band data are provided by OGLE (http://bulge.princeton.edu/\~{}ogle/ogle2/huchra.html).}
\tablenotetext{b}{The X-ray flux ratios relative to image A are estimated from 2--8 keV in order to avoid the complication of the differential absorption in the soft energy band}

\end{deluxetable}

\clearpage

\begin{deluxetable}{cccccc}
\tabletypesize{\scriptsize}
\tablecolumns{6}
\tablewidth{0pt}
\tablecaption{Kolmogorov-Smirnov Test of Variability for \cross \label{tab:ks}}
\tablehead{
\colhead{Epoch} & 
\multicolumn{5}{c}{Chance Probability\tablenotemark{a}} \\
\cline{2-6}
\colhead{} &
\colhead{Total} &
\colhead{A} & 
\colhead{B} & 
\colhead{C} & 
\colhead{D}
}
\startdata
I  & $0.11$ & $0.03$ & $0.80$ & $0.95$ & $0.08$ \\
II & $0.97$ & $0.70$ & $0.15$ & $0.86$ & $0.49$ \\
\enddata

\tablenotetext{a}{The probability that the tested light-curve is drawn from a constant distribution.}

\end{deluxetable}

\clearpage

\clearpage

\clearpage

\clearpage

\clearpage

\clearpage

\end{document}